# Twisting Light, Steering Spins: Gold Nanoparticle Magnetization via Inverse Faraday and Orbital Angular Momentum

Xingyu Yang[1], Chantal Harreau[1] and Mathieu Mivelle[1] *

[1]Sorbonne Université, CNRS, Institut des NanoSciences de Paris, INSP, F-75005 Paris, France

*Corresponding authors:

mathieu.mivelle@sorbonne-universite.fr

**Abstract**

We present a new approach to controlling magnetization in gold nanoparticles using the Inverse Faraday Effect combined with Laguerre-Gauss beams carrying orbital angular momentum. By tailoring the tilt of isophase planes, we induce drift photocurrents that generate magnetic fields tilted by up to 25° relative to the beam axis. The magnetic orientation can be reversed by switching polarization chirality or the orbital angular momentum sign, and it can be rotated azimuthally by repositioning the particle, accessing any angle over 2π steradians. This unprecedented level of control extends all-optical magnetization to three-dimensional orientations, potentially at ultrafast timescales given the near-instantaneous nature of the Inverse Faraday Effect. Our results pave the way for advanced spin-based applications, from triggering spin waves in magnetic materials to designing next-generation magnetic memory and logic devices.

**Keywords:** All-optical magnetization, Structured light, Inverse Faraday effect, Nanophotonics, Light-matter interactions, Plasmonics, Spin manipulation

## 1. Introduction

Over the past few decades, the quest to control magnetic order at ultrafast timescales has emerged as a focal point of both fundamental research and technological development, notably in the context of data storage.[1, 2] The field of ultrafast magnetism was inaugurated by Beaurepaire et al., who demonstrated that pulsed laser irradiation could induce demagnetization in metals on remarkably short timescales.[3] Building on this milestone, Kimel et al. showed that circularly polarized femtosecond laser pulses can deterministically manipulate spin dynamics via a helicity-dependent, non-thermal mechanism.[4] Subsequent studies by Stanciu et al. and Lambert et al. extended all-optical, helicity-dependent switching to ferrimagnetic and ferromagnetic materials, respectively, indicating a broadly applicable light–matter interaction.[5, 6] While the inverse Faraday effect (IFE) and magnetic circular dichroism have been identified as the primary explanatory frameworks for these observations, their relative influence remains an open question.

Spurred by these breakthroughs, researchers have continually pushed the limits of magnetization control toward smaller spatial scales and faster time domains. One promising direction involves harnessing near-instantaneous nonlinear processes in non-magnetic metallic materials—especially plasmonic metals—to drive electrons into continuous motion and thereby induce a magnetic field. This mechanism underlies the IFE in non-magnetic systems.

The IFE is a magneto-optical process that enables purely optical magnetization of matter.[7-11] This effect arises from nonlinear optical forces acting on electrons, causing drift photocurrents $J_d$ that can be described using plasma physics formalism.[12, 13] In metals, these currents follow the expression:[14, 15]

$$J_d = \frac{1}{2en} Re\left( \left( -\frac{\nabla \bullet (\sigma_\omega E)}{i\omega} \right) \bullet (\sigma_\omega E)^* \right) \qquad (1)$$

where e is the electron charge, n is the equilibrium electron density, $\sigma_\omega$ is the dynamic conductivity, and **E** represents the optical electric field. Because $J_d$ on both the electric field and its local divergence, nanophotonic and nanoplasmonic architectures—known for their ability to mold optical fields at the subwavelength level—are particularly effective for amplifying the IFE. According to the Biot–Savart law,[16-21]

$$B = \frac{\mu_0}{4\pi} \iiint \frac{J_d \times r}{|r|^3} dV \qquad (2)$$

Where $\mu_0$ is the vacuum magnetic permeability, dV is the differential volume and r is the vector from dV to the observation point, these drift currents can give rise to strong, stationary magnetic fields **B**.

Although first identified in the 1960s,[7-9] the inverse Faraday effect has garnered renewed attention with the advent of nanophotonics and ultrafast optics,[17-29] which have unlocked new avenues for manipulating magnetic processes at unprecedented temporal and spatial scales.[1, 2, 30] By finely tuning optical field gradients, intensities, and polarization, researchers have demonstrated a diverse range of magneto-optical phenomena: from generating intense, ultrafast, and localized B-fields,[19, 21, 23] to magnetizing materials with linearly polarized light,[27] imparting chirality to the magneto-optical effect,[24, 25] and steering drift currents at the nanoscale.[31] Overall, these developments highlight the expanding potential of the IFE as a powerful means of achieving ultrafast, nanoscale control over magnetic order. Nonetheless, the orientation of the induced magnetization often remains dictated by both the nanostructure and the propagation direction of the incident light, typically resulting in a magnetization that is either collinear[17] with this propagation axis or, more rarely, perpendicular to it.[31]

To further advance nanoscale plasmonic magnetization control via the IFE, we propose coupling a gold nanoparticle to a light beam carrying orbital angular momentum (OAM) to manipulate the orientation of the induced magnetization in three-dimensional space (See supplement information). Specifically, by generating a Laguerre-Gauss (LG) beam whose isophase planes are tilted by an angle set by the beam's OAM order, we tilt the plane in which drift currents are generated inside the gold nanoparticle, thereby producing a magnetic field tilted by the same angular offset. We demonstrate that this magnetization angle can be tuned by adjusting either the OAM order of the LG beam or the radial position of the nanoparticle within the beam, enabling control of the magnetic orientation over a range from 0° to 25° from the normal. Furthermore, by varying the azimuthal position of the nanoparticle in the Laguerre-Gauss beam, the magnetization orientation can span any angle θ over 2π steradians. This unprecedented level of control at the nanoscale—potentially on ultrafast timescales—opens up new possibilities for controlling magnetic materials and thin films, for example in generating spin waves whose initial spin orientation is set by the nanoparticle's magnetization. This breakthrough thus paves the way toward next-generation magnetic memory and logic devices for data writing and processing.

**2. Results:**

In most cases, the magnetization induced by the Inverse Faraday Effect in a material is collinear with the propagation direction of the incident light.[17, 21] This also holds true for a gold nanoparticle excited by a plane wave or a Gaussian beam whose isophase planes are perpendicular to the propagation direction of the incident wave (Figure 1a). In such a scenario, the drift currents responsible for the magnetization are generated within the wave's isophase planes, producing a magnetization aligned with the wave vector. However, exciting the same nanoparticle with a LG beam carrying OAM means that the isophase planes are no longer perpendicular to the propagation direction, raising the question of whether the resulting drift currents remain confined to those isophase planes. If they do, the ensuing magnetization would be tilted (Figure 1b).

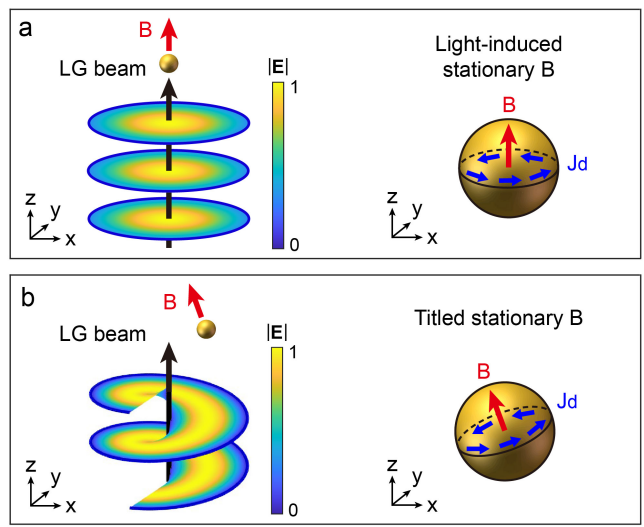

**Figure 1**. a) Inverse Faraday Effect magnetization of a gold nanoparticle excited by a Laguerre-Gauss beam whose isophase planes are perpendicular to the propagation direction of the incident wave.[17] b) Magnetization attainable through IFE when these isophase planes, instead of being perpendicular, are tilted by an angle θ with respect to the propagation direction.

To achieve this, we carried out numerical simulations summarized in Figure 2a. A gold nanoparticle with a diameter of 100 nm is placed in a LG beam carrying OAM, where the intensity maximum is located at r=1500 nm from the center of the beam, for a wavelength of 545 nm and a right-handed circular polarization (RHCP). The nanoparticle can then be moved radially (r) or azimuthally (φ) within the same LG beam (Figure 2a). Figure 2b shows the distribution of the optical phase in a YZ plane passing through the intensity maximum along X, at a distance of 1500 nm from the beam center, as indicated in Figure 2a. As can be seen, the isophases exhibit a slope of θ=13,1 degrees with respect to the Y axis. Once placed in the phase plane from Figure 2b, the interaction between the light and the nanoparticle produces the electric field amplitude distributions shown in Figures 2c and 2d. Specifically, Figure 2c depicts the distribution in the XZ plane, and Figure 2d shows it in the YZ plane. In the XZ plane, the electric field is primarily oriented along X, featuring two optical hot spots on either side of the nanoparticle, as expected for a wave propagating along Z with a transverse electric field. In contrast, the distribution in the YZ plane does not exhibit the same pattern. Indeed, the hot spots in this plane are tilted relative to the Y and Z axes by an angle matching that of the isophase in Figure 2b. From these electric field distributions and Equation 1, we can calculate the associated drift photocurrents in the metal nanoparticle. Figures 2e and 2f show these drift currents $\mathbf{J_d}$ in an XY plane at the center of the particle and in 3D within a quarter-hemisphere, respectively. As can be seen, these currents are not perpendicular to the Z axis—i.e., not strictly aligned perpendicular with the wave's propagation direction—but instead exhibit a twisted orientation that depends on the specific position within the nanoparticle.

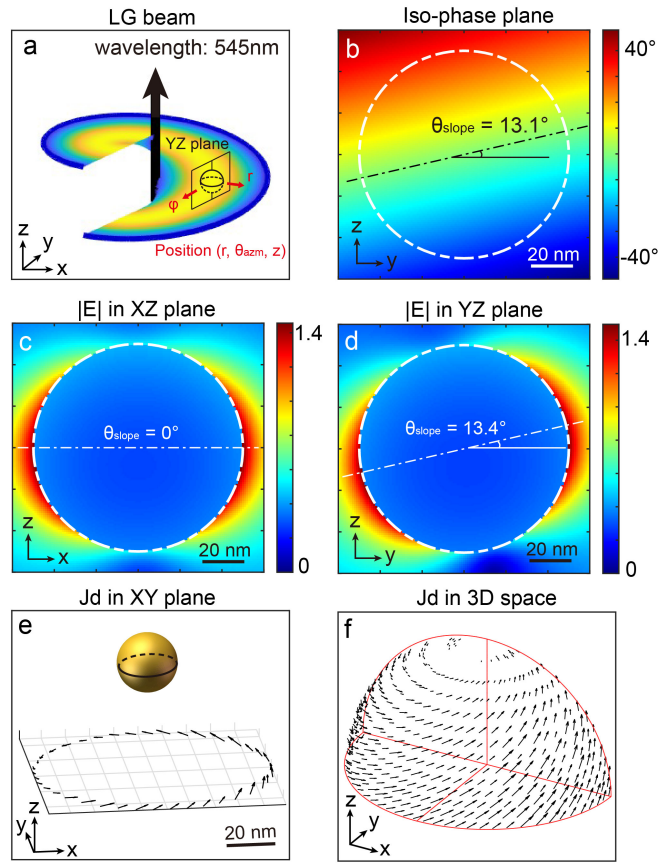

**Figure 2**. IFE-induced magnetization of a gold nanoparticle in an LG beam carrying an OAM of order 4. a) Simulation schematic: a gold nanoparticle with a diameter of 100 nm is placed 1500 nm from the beam center, which is right-circularly polarized and provides maximum intensity at a wavelength of 545 nm. b) Phase map experienced by the nanoparticle in a YZ plane, highlighting a tilt of about 13°. c) Electric field amplitude distribution near the nanoparticle in an XZ plane, and d) in a YZ plane. e) Spatial distribution of drift currents in the nanoparticle's skin depth within an XY plane at its center, and f) in 3 dimensions within a quarter hemisphere.

Using the drift currents calculated and shown in Figures 2e,f, together with Equation 2, it is possible to derive the associated magnetic field (shown in SI). Figure 3 illustrates the orientation of this magnetic field for various OAM excitations. Figures 3a,b present this orientation for an OAM of order 4 with right- and left-circular polarization, respectively, while Figures 3c,d show the orientation of **B** for the same polarizations but with an OAM of order -4.

First, as can be observed, the magnetic field direction is tilted in the YZ plane by an angle θ relative to the Z axis. This finding is particularly noteworthy since, to the best of our knowledge, it constitutes the first demonstration of external manipulation of magnetization direction via the IFE. In addition, as predicted by the IFE, changing the polarization from right-circular to left-circular flips the magnetization orientation by 180°. Likewise, transitioning from an OAM of order 4 to -4 changes the tilt angle from θ to -θ.

Finally, as shown in Figure 3e, rotating the nanoparticle's position around the center of the LG beam at a fixed radial distance enables the magnetic field orientation to rotate through 2pi steradians about the Z axis, all while maintaining the same tilt angle θ.

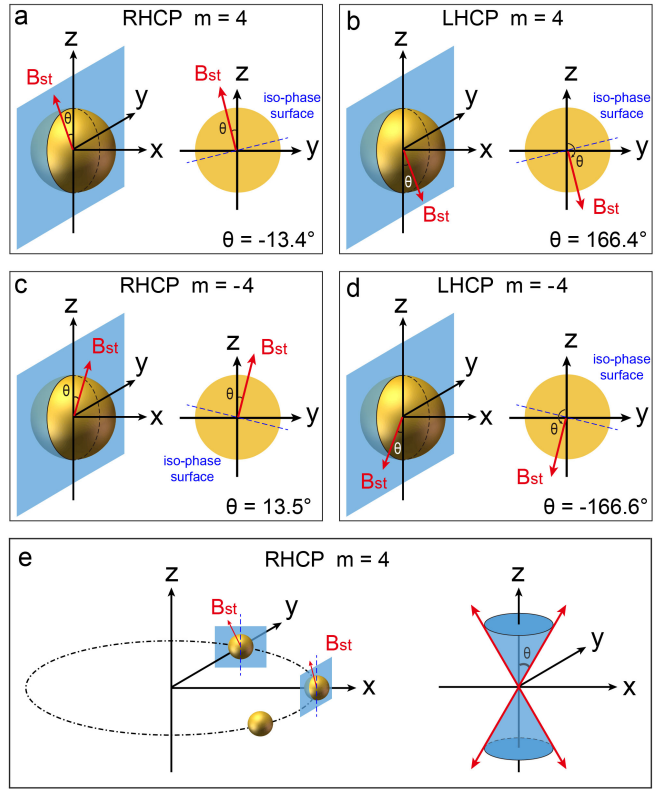

**Figure 3**. Magnetic field orientation as a function of the LG beam parameters. a,b) Orientation of the **B** field for an OAM of order 4 and c,d) of order -4, with a,c) right-circular polarization and b,d) left-circular polarization. These four configurations yield four distinct magnetization orientations: θ = -13.4°, 13.5°, 166.4°, and -166.6°. e) Magnetic field orientation for an OAM of order 4 and right-circular polarization, shown for various azimuthal positions of the nanoparticle within the LG beam. This azimuthal adjustment allows the magnetic field orientation to rotate around the Z axis.

In order to further manipulate the direction of the optically generated **B** field, Figure 4 examines the influence of the OAM number, the nanoparticle's position within the beam, and the nanoparticle's size. To that end, Figure 4a illustrates various LG beams carrying different OAM numbers, highlighting the helical phase structures characteristic of these beams. Figure 4b shows the tilt θ of the **B** field orientation induced by the IFE when the nanoparticle is placed at the intensity maximum (r = 1500nm) of an LG beam with OAM numbers ranging from 0 to 8. As expected, for an OAM number of 0, the tilt is zero. However, as the OAM number increases, the tilt θ grows linearly in discrete increments, reaching about 25 degrees. By contrast, as shown in Figure 4c, increasing the nanoparticle's radius does not affect this tilt. For OAM numbers of 3, 4, or 5 and radii between 25 and 50 nm, the magnetic deviation angles remain unchanged.

Interestingly, a way to bridge the gap between the discrete tilt values shown in Figure 4b for each OAM number is to shift the nanoparticle's position within the LG beam. Indeed, as illustrated in Figure 4d, for an OAM number of 4, centering the nanoparticle at the beam's maximum electric field yields, as previously noted, a magnetization tilt of θ=13.4°. However, by moving the nanoparticle radially in the LG beam, it becomes possible to tune the magnetic deviation by approximately ±3°, allowing one to access the deviation values corresponding to OAM orders of ±1. Ultimately, by varying the OAM number or the nanoparticle's position, it is feasible to continuously adjust the tilt θ from 0° to 25° (Figure 4b) over 2pi steradians.

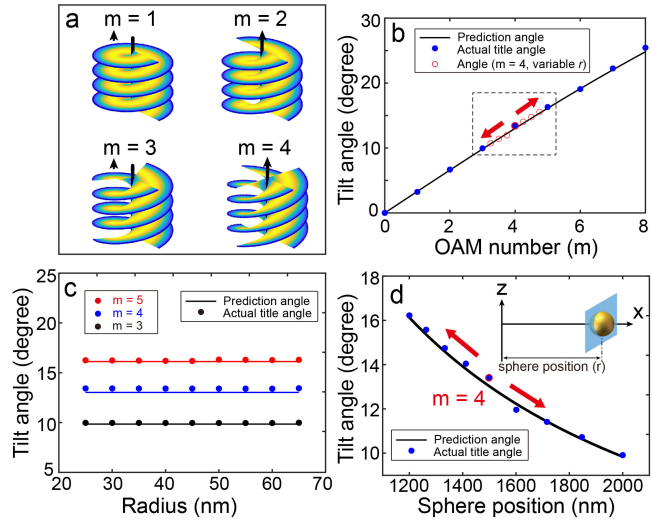

**Figure 4**. Control of the magnetic field **B** tilt by the OAM order and the nanoparticle's position. a) Illustration of LG beams characterized by different OAM orders, highlighting their helical phase structure. b) Evolution of the **B** field tilt angle θ as a function of the OAM order, for a nanoparticle placed at the beam's intensity maximum: the angle is zero for an OAM order of 0 and increases linearly to about 25° for an OAM order of 8. The theoretical value (solid black line) corresponds to the tilt angle of the isophase planes, which can be directly predicted based on the position r and the OAM number (as detailed in the Supplementary Information). c) Influence of the nanoparticle's radius on this angle for fixed OAM orders of 3, 4, and 5. d) The radial displacement of the nanoparticle within the LG beam (here for an OAM order of 4) allows fine-tuning of the tilt angle by approximately ±3° around its central value (about 13.4°), thus offering nearly continuous control of the **B** field orientation from 0° to 25° by combining the OAM order with the particle's position in the beam.

In conclusion, in this theoretical article, we have demonstrated that by using a simple gold nanoparticle and an LG beam carrying a certain OAM order, it is possible to generate a magnetization that is tilted by a nonzero angle with respect to the direction of light propagation. We have also shown that the magnetization orientation can be chosen to span 2pi steradians and reversed by switching the incident polarization between right- and left-circular or by changing the sign of the OAM number. Finally, we have demonstrated that increasing or decreasing the OAM number or shifting the nanoparticle's radial position within the LG beam allows one to continuously control the magnetic deflection angle from 0° to 25°. Overall, this level of control enables a fine-tuning of the magnetization orientation within a solid angle around the wave propagation axis, ranging from 0 to ±θ and from pi to ±(pi+θ). To the best of our knowledge, this is the first time such control has been proposed in the literature. The results presented here go beyond a mere academic demonstration, opening the way to highly precise control of magnetic processes on potentially ultrafast timescales, due to the nearly instantaneous nature of the IFE. One of the applications we envision involves launching spin waves in a magnetic material, where the initial spin orientation is determined by the nanoparticle's magnetization, or the fine manipulation of magnetization reversal in magnetic alloy.[32]

# Bibliography

(1) Bossini, D.; Belotelov, V. I.; Zvezdin, A. K.; Kalish, A. N.; Kimel, A. V. Magnetoplasmonics and femtosecond optomagnetism at the nanoscale. *Acs Photonics* **2016**, *3* (8), 1385-1400.

(2) Kimel, A. V.; Li, M. Writing magnetic memory with ultrashort light pulses. *Nature Reviews Materials* **2019**, *4* (3), 189-200.

(3) Beaurepaire, E.; Merle, J.-C.; Daunois, A.; Bigot, J.-Y. Ultrafast spin dynamics in ferromagnetic nickel. *Phys. Rev. Lett.* **1996**, *76* (22), 4250.

(4) Kimel, A.; Kirilyuk, A.; Usachev, P.; Pisarev, R.; Balbashov, A.; Rasing, T. Ultrafast non-thermal control of magnetization by instantaneous photomagnetic pulses. *Nature* **2005**, *435* (7042), 655-657.

(5) Stanciu, C.; Kimel, A.; Hansteen, F.; Tsukamoto, A.; Itoh, A.; Kirilyuk, A.; Rasing, T. Ultrafast spin dynamics across compensation points in ferrimagnetic GdFeCo: The role of angular momentum compensation. *Phys. Rev. B* **2006**, *73* (22), 220402.

(6) Lambert, C.-H.; Mangin, S.; Varaprasad, B. C. S.; Takahashi, Y.; Hehn, M.; Cinchetti, M.; Malinowski, G.; Hono, K.; Fainman, Y.; Aeschlimann, M. All-optical control of ferromagnetic thin films and nanostructures. *Science* **2014**, *345* (6202), 1337-1340.

(7) Pitaevskii, L. Electric forces in a transparent dispersive medium. *Sov. Phys. JETP* **1961**, *12* (5), 1008-1013.

(8) Van der Ziel, J.; Pershan, P. S.; Malmstrom, L. Optically-induced magnetization resulting from the inverse Faraday effect. *Phys. Rev. Lett.* **1965**, *15* (5), 190.

(9) Pershan, P.; Van der Ziel, J.; Malmstrom, L. Theoretical discussion of the inverse Faraday effect, Raman scattering, and related phenomena. *Phys. Rev.* **1966**, *143* (2), 574.

(10) Battiato, M.; Barbalinardo, G.; Oppeneer, P. M. Quantum theory of the inverse Faraday effect. *Phys. Rev. B* **2014**, *89* (1), 014413.

(11) Berritta, M.; Mondal, R.; Carva, K.; Oppeneer, P. M. Ab initio theory of coherent laser-induced magnetization in metals. *Phys. Rev. Lett.* **2016**, *117* (13), 137203.

(12) Karpman, V.; Shagalov, A. The ponderomotive force of a high-frequency electromagnetic field in a cold magnetized plasma. *Journal of Plasma Physics* **1982**, *27* (2), 215-224.

(13) Hora, H. *Laser Plasma Physics: forces and the nonlinearity principle*; Spie Press, 2000.

(14) Hertel, R. Theory of the inverse Faraday effect in metals. *J. Magn. Magn. Mater.* **2006**, *303* (1), L1-L4.

(15) Hertel, R.; Fähnle, M. Macroscopic drift current in the inverse Faraday effect. *Phys. Rev. B* **2015**, *91* (2), 020411.

(16) Smolyaninov, I. I.; Davis, C. C.; Smolyaninova, V. N.; Schaefer, D.; Elliott, J.; Zayats, A. V. Plasmon-induced magnetization of metallic nanostructures. *Phys. Rev. B* **2005**, *71* (3), 035425.

(17) Nadarajah, A.; Sheldon, M. T. Optoelectronic phenomena in gold metal nanostructures due to the inverse Faraday effect. *Opt. Express* **2017**, *25* (11), 12753-12764.

(18) Hurst, J.; Oppeneer, P. M.; Manfredi, G.; Hervieux, P.-A. Magnetic moment generation in small gold nanoparticles via the plasmonic inverse Faraday effect. *Phys. Rev. B* **2018**, *98* (13), 134439.

(19) Cheng, O. H.-C.; Son, D. H.; Sheldon, M. Light-induced magnetism in plasmonic gold nanoparticles. *Nat. Photonics* **2020**, *14* (6), 365-368.

(20) Sinha-Roy, R.; Hurst, J.; Manfredi, G.; Hervieux, P.-A. Driving orbital magnetism in metallic nanoparticles through circularly polarized light: A real-time tddft study. *ACS photonics* **2020**, *7* (9), 2429-2439.

(21) Yang, X.; Mou, Y.; Gallas, B.; Maitre, A.; Coolen, L.; Mivelle, M. Tesla-Range Femtosecond Pulses of Stationary Magnetic Field, Optically Generated at the Nanoscale in a Plasmonic Antenna. *ACS Nano* **2021**.
(22) Popova, D.; Bringer, A.; Blügel, S. Theory of the inverse Faraday effect in view of ultrafast magnetization experiments. *Phys. Rev. B* **2011**, *84* (21), 214421.
(23) Cheng, O. H.-C.; Zhao, B.; Brawley, Z.; Son, D. H.; Sheldon, M. T. Active Tuning of Plasmon Damping via Light Induced Magnetism. *Nano. Lett.* **2022**, *22* (13), 5120-5126.
(24) Mou, Y.; Yang, X.; Gallas, B.; Mivelle, M. A Reversed Inverse Faraday Effect. *Advanced Materials Technologies* **2023**, *n/a* (n/a), 2300770. DOI: https://doi.org/10.1002/admt.202300770.
(25) Mou, Y.; Yang, X.; Gallas, B.; Mivelle, M. A chiral inverse Faraday effect mediated by an inversely designed plasmonic antenna. *Nanophotonics* **2023**, *12* (12), 2115-2120.
(26) Ortiz, V. H.; Mishra, S. B.; Vuong, L.; Coh, S.; Wilson, R. B. Specular inverse Faraday effect in transition metals. *Physical Review Materials* **2023**, *7* (12), 125202.
(27) Yang, X.; Mou, Y.; Zapata, R.; Reynier, B.; Gallas, B.; Mivelle, M. An inverse Faraday effect generated by linearly polarized light through a plasmonic nano-antenna. *Nanophotonics* **2023**, (0).
(28) González-Alcalde, A. K.; Shi, X.; Ortiz, V. H.; Feng, J.; Wilson, R. B.; Vuong, L. T. Enhanced inverse Faraday effect and time-dependent thermo-transmission in gold nanodisks. *Nanophotonics* **2024**, (0).
(29) Lian, D.; Yang, Y.; Manfredi, G.; Hervieux, P.-A.; Sinha-Roy, R. Orbital magnetism through inverse Faraday effect in metal clusters. *Nanophotonics* **2024**, *13* (23), 4291-4302.
(30) Parchenko, S.; Hofhuis, K.; Larsson, A. Å.; Kapaklis, V.; Scagnoli, V.; Heyderman, L. J.; Kleibert, A. Plasmon-Enhanced Optical Control of Magnetism at the Nanoscale via the Inverse Faraday Effect. *Advanced Photonics Research* **2025**, *6* (1), 2400083.
(31) Mou, Y.; Yang, X.; Vega, M.; Zapata, R.; Gallas, B.; Bryche, J.-F.; Bouhelier, A.; Mivelle, M. Femtosecond drift photocurrents generated by an inversely designed plasmonic antenna. *Nano. Lett.* **2024**.
(32) Stenning, K. D.; Xiao, X.; Holder, H. H.; Gartside, J. C.; Vanstone, A.; Kennedy, O. W.; Oulton, R. F.; Branford, W. R. Low-power continuous-wave all-optical magnetic switching in ferromagnetic nanoarrays. *Cell Reports Physical Science* **2023**, *4* (3).

**Supplementary information**

# Twisting Light, Steering Spins: Gold Nanoparticle Magnetization via Inverse Faraday and Orbital Angular Momentum

Xingyu Yang[1], Chantal Harreau[1] and Mathieu Mivelle[1] *

[1]Sorbonne Université, CNRS, Institut des NanoSciences de Paris, INSP, F-75005 Paris, France

*Corresponding authors:

mathieu.mivelle@sorbonne-universite.fr

## 2. Laguerre–Gaussian Beam

In 1992, it was discovered that laser light with a Laguerre–Gaussian (LG) distribution carries a well-defined orbital angular momentum (OAM) [1]. The beam in our study is a typical LG beam described analytically in cylindrical coordinates [1]. Locally, it is circularly polarized to genrate an inverse Faraday effect (IFE). In the plane perpendicular to the beam's propagation, the OAM is associated with the phase factor $\exp(-im\phi)$, where m is the OAM index and $\phi$ is the azimuthal angle.

We place a gold nanosphere at the beam waist plane, at a radial distance r = 1500

r=1500 nm from the beam's hollow core. To ensure that the LG beam's intensity maximum aligns with the particle's position, the beam waist is adjusted according to different m values. This configuration allows us to systematically study the influence of OAM on the induced magnetization.

## 3. Numerical Simulation Parameters

All simulations were conducted in the frequency domain using COMSOL's Wave Optics Module. The Laguerre–Gaussian (LG) beam was analytically defined as a background scattered field in accordance with Ref. [1], serving as the incident illumination. Perfectly matched layers (PMLs) were implemented around the computational domain to emulate unbounded space.

To accurately capture nanoscale light–matter interactions, a fine mesh (mesh size < 3 nm) was applied to the gold nanosphere. The surrounding air region was meshed at $< \lambda/6$, and a boundary layer mesh was also included between the particle and the air. Simulation results were then exported on a $1 \times 1 \times 1$ nm³ cubic grid for subsequent calculations of the inverse Faraday effect and the stationary magnetic field.

## 4. Predicted Tilt Angle of the Isophase Plane

Laguerre–Gaussian (LG) beams exhibit a spiral isophase plane in three-dimensional space. If we "unwrap" this spiral along the beam axis (z-axis) over a $2\pi$ azimuthal range, the isophase surfaces form a right triangle. The tilt angle of these surfaces depends on the orbital angular momentum (OAM) index (m), the light wavelength $\lambda=545$nm, and the unwrapping radius r=1500nm. This relationship can be expressed as:

$$\theta = \arctan\left(\frac{m\lambda}{2\pi r}\right) \quad (S1)$$

Equation (S1) provides the theoretical prediction for the isophase plane tilt angle plotted in Figure 4.

## 5. Light-induced stationary B field

The spin density represents the rotation of the electric field in its local plane and is defined as the cross product of the electric field with its complex conjugate (Equation S2):

$$\mathbf{s} = \frac{1}{E_0^2}\mathrm{Im}(\mathbf{E}^* \times \mathbf{E}) \quad (S2)$$

where E0 is the amplitude of the incident electric field. Within the framework of the inverse Faraday effect (IFE), spin density signals the presence of induced drift photocurrents. Figures S1(a–c) illustrate the amplitude of the spin density across three representative cross-sections, with Figure S1(c) clearly revealing a tilt in the YZ-plane. Consequently, the gold nanosphere exhibits an internally tilted magnetization generated by the IFE.

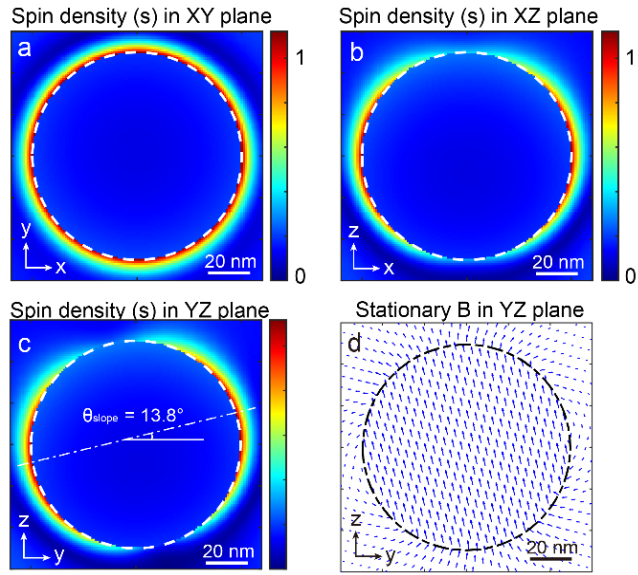

**Figure S1**. a–c) Spin density amplitude in the a) XY, b) XZ, and c) YZ cross-sections. The YZ plane shows a tilt of approximately 13.8°, matching the tilt angle of the stationary **B** field. d) IFE-induced stationary **B** field in the YZ plane, revealing a tilted magnetization inside the nanosphere.

**Reference**

[1] Allen, Les, et al. "Orbital angular momentum of light and the transformation of Laguerre-Gaussian laser modes." Physical review A 45.11 (1992): 8185.